\documentclass[times, review, 10pt]{elsarticle}

\usepackage{amssymb}
\usepackage{amsmath}
\usepackage{booktabs}
\usepackage{multirow}
\usepackage[table,xcdraw]{xcolor}
\usepackage[numbers]{natbib}
\usepackage{adjustbox}
\usepackage{relsize}

\usepackage{utfsym}
\usepackage{times}
\usepackage{soul}
\usepackage{url}
\usepackage[utf8]{inputenc}
\usepackage{graphicx}
\usepackage{amsmath}
\usepackage{amsthm}
\usepackage{booktabs}
\usepackage[switch]{lineno}
\usepackage{blindtext}
\usepackage{makecell}
\usepackage{pifont}
\usepackage{amssymb}
\usepackage{color}
\usepackage{tikz}
\usepackage[edges]{forest}
\usepackage{multirow}
\definecolor{hidden-draw}{RGB}{205, 44, 36}
\definecolor{hidden-blue}{RGB}{194,232,247}
\definecolor{hidden-orange}{RGB}{243,202,120}
\definecolor{hidden-yellow}{RGB}{242,244,193}
\usepackage{cleveref}
\usepackage{tcolorbox}

\usepackage{xcolor}
\usepackage{ifthen}
\usepackage[normalem]{ulem}

\definecolor{myorange}{HTML}{FEAE03}
\definecolor{myturquois}{HTML}{01AB8F}
\definecolor{mypink}{HTML}{D31876}

\definecolor{brightred}{HTML}{E55347} 
\definecolor{orange}{HTML}{FF8C00} 
\definecolor{yellowgreen}{HTML}{6B8E23} 
\definecolor{green}{HTML}{228B22} 
\usepackage{tikz}
\usepackage[edges]{forest}
\usetikzlibrary{shapes, arrows.meta, positioning}
\usepackage{graphicx}
\usepackage{forest}
\usetikzlibrary{trees,positioning,shapes,shadows,arrows.meta}

\definecolor{myblue}{RGB}{159, 192, 230}
\definecolor{myblueline}{RGB}{87, 127, 185}
\definecolor{bluelight1}{RGB}{185, 211, 237}
\definecolor{bluelight2}{RGB}{213, 222, 239}
\definecolor{mygreen}{RGB}{168, 209, 201}
\definecolor{greenlight}{RGB}{220, 235, 234}
\definecolor{hidden-draw}{RGB}{177, 177, 177}
\definecolor{mygray}{RGB}{185, 185, 185}
\definecolor{lightcoral}{rgb}{0.94, 0.5, 0.5}
\definecolor{lightgreen}{rgb}{0.56, 0.93, 0.56}
\definecolor{harvestgold}{rgb}{0.98, 0.85, 0.40}
\definecolor{brightlavender}{rgb}{0.75, 0.58, 0.89}
\definecolor{capri}{rgb}{0.0, 0.75, 1.0}
\definecolor{carminepink}{rgb}{0.92, 0.3, 0.26}
\definecolor{celadon}{rgb}{0.67, 0.88, 0.69}
\definecolor{darkpastelgreen}{rgb}{0.01, 0.75, 0.24}
\definecolor{lightgoldenrodyellow}{rgb}{0.98, 0.98, 0.82}
\definecolor{jonquil}{rgb}{0.98, 0.85, 0.37}
\definecolor{lightkhaki}{rgb}{0.94, 0.9, 0.55}
\definecolor{lemonchiffon}{rgb}{1.0, 0.98, 0.8}
\definecolor{schoolbusyellow}{rgb}{1.0, 0.85, 0.0}

\begin{document}

\begin{frontmatter}

\title{A Survey of LLM Alignment: Instruction Understanding, Intention Reasoning, and Reliable Generation} 

\author[1]{Zongyu Chang*}
\author[2]{Feihong Lu*}
\author[6]{Ziqin Zhu}
\author[1]{Qian Li}
\author[2]{Cheng Ji}
\author[2]{Tao Yang}
\author[1]{Zhuo Chen}
\author[2]{Hao Peng}
\author[5]{Yang Liu}
\author[3]{Ruifeng Xu}
\author[4]{Yangqiu Song}
\author[2]{Jianxin Li}
\author[1]{Shangguang Wang}
\fntext[1]{* These authors contributed equally to this work.}

\affiliation[1]{organization={Beijing University of Posts and Telecommunications},
            city={Beijing},
            postcode={100876}, 
            country={China}}
\affiliation[2]
{organization={Beihang University},
            city={Beijing},
            postcode={100191}, 
            country={China}}
\affiliation[3]
{organization={Harbin Institute of Technology},
            city={Harbin},
            postcode={150081}, 
            country={China}}
\affiliation[4]
{organization={Hong Kong University of Science and Technology},
            city={Hong Kong},
            country={China}}
\affiliation[5]
{organization={Chinese Academy of Sciences},
            city={Beijing},
            postcode={230026}, 
            country={China}}
\affiliation[6]
{organization={The University of Auckland},
            city={Auckland},
            country={New Zealand}}
\begin{abstract}
Large language models have demonstrated exceptional capabilities in understanding and generation. However, in real-world scenarios, users’ natural language expressions are often inherently fuzzy, ambiguous, and uncertain, leading to challenges such as vagueness, polysemy, and contextual ambiguity. 
This paper focuses on three challenges in LLM-based text generation tasks: instruction understanding, intention reasoning, and reliable dialog generation. 
Regarding human complex instruction, LLMs have deficiencies in understanding long contexts and instructions in multi-round conversations. For intention reasoning, LLMs may have inconsistent command reasoning, difficulty reasoning about commands containing incorrect information, difficulty understanding user ambiguous language commands, and a weak understanding of user intention in commands. Besides, In terms of Reliable Dialog Generation, LLMs may have unstable generated content and unethical generation. To this end, we classify and analyze the performance of LLMs in challenging scenarios and conduct a comprehensive evaluation of existing solutions.
Furthermore, we introduce benchmarks and categorize them based on the aforementioned three core challenges. 
Finally, we explore potential directions for future research to enhance the reliability and adaptability of LLMs in real-world applications.
\end{abstract}

\begin{keyword}
Instruction Understanding, Intention Reasoning, Reliable Dialog Generation
\end{keyword}

\end{frontmatter}

\section{Introduction}
Rapid advancements with the development of large language models (LLMs) have been experienced in the field of artificial intelligence. 
These models, built upon massive amounts of data and extensive computing resources, have shown impressive capabilities in understanding and generating human language. 
Recent advancements in LLMs, including the use of scaling laws~\cite{kaplan2020scaling}, supervised fine-tuning (SFT)~\cite{wu2021recursively}, and reinforcement learning with human feedback (RLHF)~\cite{ouyang2022training}, have propelled these models to new heights. Researchers have explored innovative strategies like chain-of-thought reasoning (COT)~\cite{wei2022chain}, aiming to enhance their performance in processing and generating accurate responses. 
However, they still have struggled when interacting with human instructions in real-world scenarios due to the fuzzy, uncertain, and complex nature of human instructions, especially when the input data is fuzzy, incomplete, or inconsistent. Despite improvements, issues such as content hallucination~\cite{DBLP:conf/emnlp/LiCZNW23} and logical misinterpretations remain prevalent. Consequently, while LLMs show promise, they are far from flawless and require further refinement to address the challenges posed by more unpredictable and complex human instructions as follows.

\begin{figure}[t]
  \setlength{\abovecaptionskip}{-0.1cm}
  \setlength{\belowcaptionskip}{-0.3cm}
  \centering
  \includegraphics[width=1
  \linewidth]{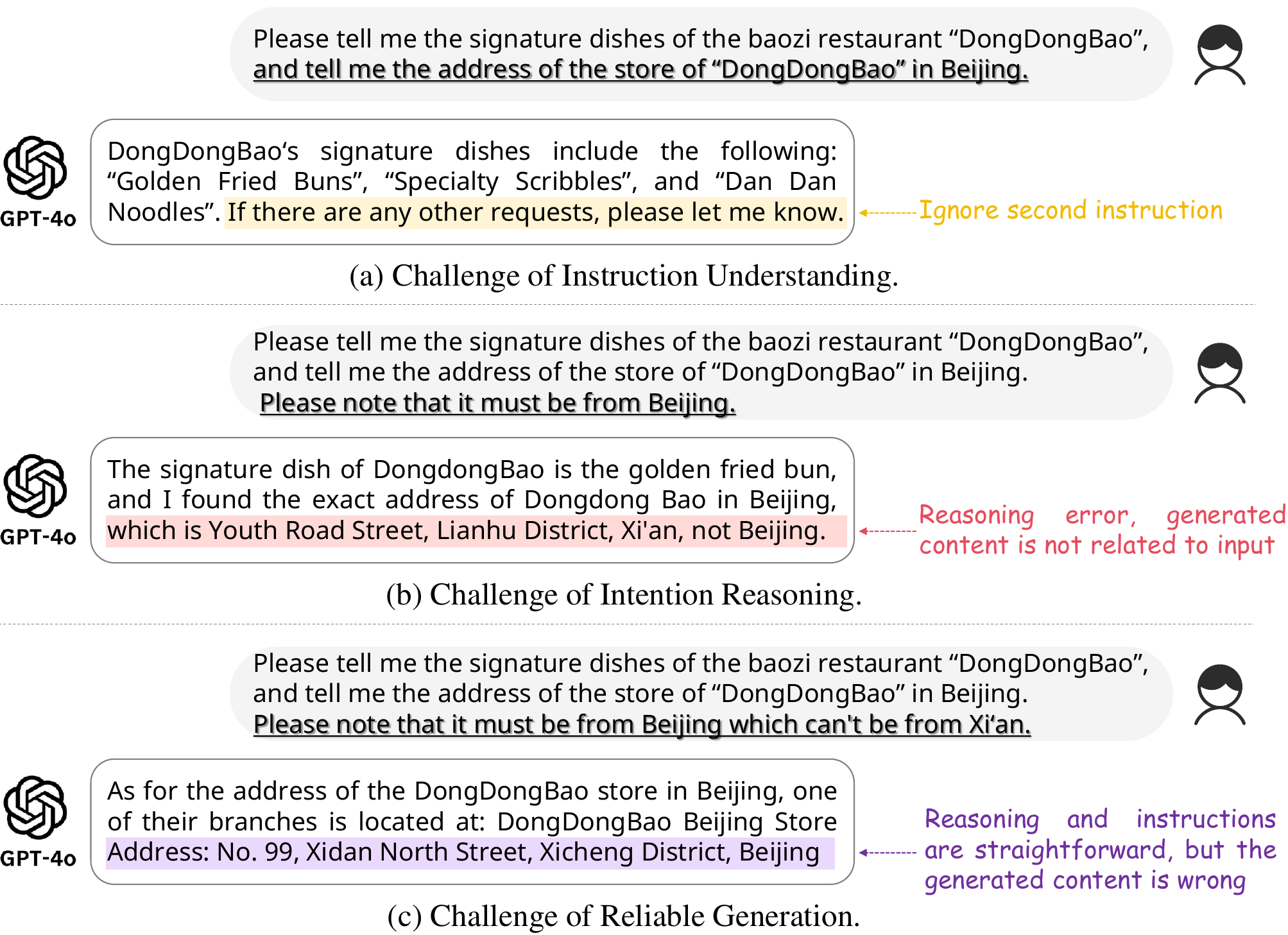}
  \caption{Example of LLMs generation.}\label{fig:intro}
\end{figure}

\begin{figure*}[t]
  \centering
  \setlength{\abovecaptionskip}{-0.1cm}
  \setlength{\belowcaptionskip}{-0.1cm}
  \includegraphics[width=1
  \linewidth]{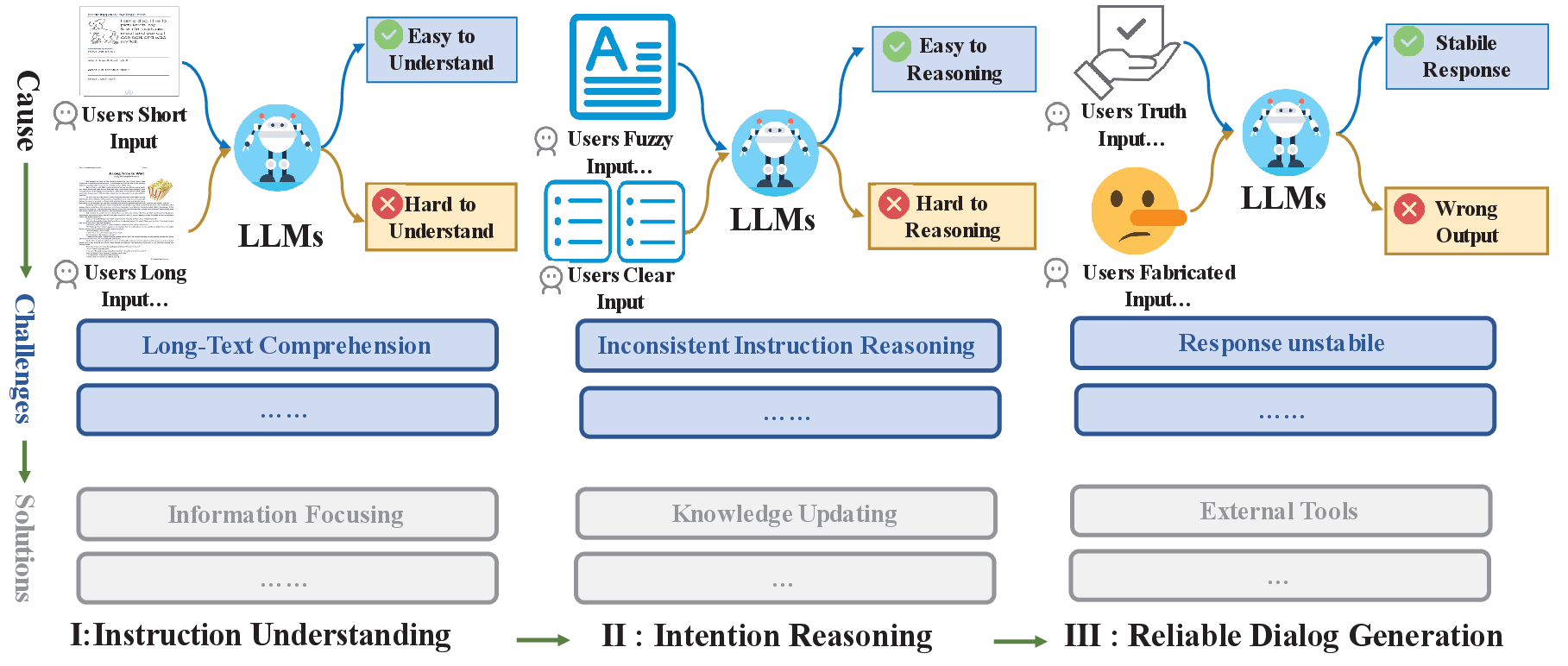}
  \caption{Unlike previous surveys on LLMs, we do not consider the alignment between LLMs and humans as an isolated process,instead, we view it as a continuous and dynamic information processing process consisting of instruction understanding, intention reasoning, and reliable dialogue generation.}\label{fig:intro1}
\end{figure*}

\subsection{Challenge of Instruction Understanding}
One of the most pressing challenges that LLMs face is instruction understanding as \Cref{fig:intro}(a) and Figure (\ref{fig:intro1}\uppercase\expandafter{\romannumeral1}), particularly when the user input involves complex or multi-step instructions. While models have improved in parsing relatively simple queries, they continue to encounter significant difficulties when dealing with long, context-rich instructions or when instructions are spread across multiple conversational turns. LLMs often fail to grasp subtle nuances or interpret implicit meanings embedded within the text, which leads to inaccurate or incomplete responses. Existing approaches to instruction understanding have introduced techniques like optimizing the model's parsing abilities~\cite{teng2024fine}, and context-aware optimization~\cite{sun2024parrot}. While these methods show promise, they often fall short when addressing the complexities and ambiguities present in instructions.

\subsection{Challenge of Intention Reasoning}
Another critical area is intention reasoning as illustrated in \Cref{fig:intro}(b) and Figure \ref{fig:intro1} (\uppercase\expandafter{\romannumeral2}), where they struggle to align the generated responses with the user’s underlying intention. Ambiguities in language, conflicting instructions, and implicit requirements often result in models producing outputs that diverge from the user's expectations. LLMs also face difficulties when instructions are inconsistent or contain incorrect information, which challenges the model’s ability to make accurate inferences. Various strategies, including retrieval-enhanced generation and fine-tuning techniques, have been proposed to enhance reasoning capabilities, enabling the models to better handle inconsistent or incomplete instructions. However, these methods often introduce new challenges related to bias and the inability to fully resolve conflicts in user input, further complicating the alignment between generated content and user expectations.

\subsection{Challenge of Reliable Dialog Generation.}
The final major challenge is the reliable dialog generation, which pertains to the accuracy, ethical considerations, and stability of the content they produce, such as \Cref{fig:intro}(c) and Figure \ref{fig:intro1} (\uppercase\expandafter{\romannumeral3}. While LLMs are generally capable of generating coherent and contextually relevant outputs, they sometimes exhibit instability, generating content that is factually incorrect, logically inconsistent, or ethically questionable. This challenge is exacerbated by the model’s inability to recognize uncertainty, which can lead to overconfident but inaccurate outputs. Recent efforts to address this issueinvolve techniques like uncertainty-aware fine-tuning and using external tools to evaluate output credibility. However, these approaches struggle to provide a comprehensive and reliable solution, especially in complex or dynamic contexts.

\tikzstyle{my-box}=[
    rectangle,
    draw=hidden-draw,
    rounded corners,
    text opacity=1,
    minimum height=1.5em,
    minimum width=5em,
    inner sep=2pt,
    align=center,
    fill opacity=.5,
]

\tikzstyle{understanding_leaf}=[my-box, minimum height=1.5em,
    fill=bluelight2!100, text=black, align=left,font=\scriptsize,
    inner xsep=2pt,
    inner ysep=4pt,
]
\tikzstyle{reasoing_leaf}=[my-box, minimum height=1.5em,
    fill=greenlight, text=black, align=left,font=\scriptsize,
    inner xsep=2pt,
    inner ysep=4pt,
]
\tikzstyle{reliable_leaf}=[my-box, minimum height=1.5em,
    fill=lightgoldenrodyellow, text=black, align=left,font=\scriptsize,
    inner xsep=2pt,
    inner ysep=4pt,
]


\begin{figure*}[!t]
    \centering
    \setlength{\abovecaptionskip}{-0.1cm}
    \resizebox{\textwidth}{!}{
        \begin{forest}
            forked edges,
            for tree={
                grow=east,
                reversed=true,
                base=left,
                font=\scriptsize,
                rectangle,
                draw=hidden-draw,
                rounded corners,
                align=left,
                minimum width=4em,
                edge+={darkgray, line width=1pt},
                inner xsep=2pt,
                inner ysep=3pt,
                ver/.style={rotate=90, child anchor=north, anchor=center}
            },
            [
                Gap Between LLMs and Human Intention, ver, color=hidden-draw, fill=mygray!30,
                text=black
                [
                    Instruction \\ Understanding (\S \ref{sec:challenge1}), color=myblue!100, fill=myblue!80, text width=6em, text=black
                    [
                        Long-text \\ Comprehension (\S \ref{sec:long}), color=bluelight1!60, fill=bluelight1!60, text width=7em, text=black
                        [
                            Information \\ Focusing, color=bluelight1!40, fill=bluelight1!40,  text width=4em, text=black [
                                1) Attention Sparsification \cite{beltagy2020longformer}{,}
                                2) Dynamically adjusting attention\cite{wu2021smart}{,}\\
                                3) Attention Optimization \cite{chen2024core}{,}
                                4) Position independent Training \cite{he2024never}
                                , understanding_leaf, text width=26.5em
                            ]
                        ]
                        [
                            Multipath \\ Optimization, color=bluelight1!40, fill=bluelight1!40,  text width=4em, text=black
                            [
                                1) SFT and RL \cite{zhang2024longreward,bai2024longalign,tang2024logo,chen2023longlora}{,} 
                                2) Retrieval-Augmented \cite{li2024retrieval,lewis2020retrieval}{,}\\
                                3) Recurrent-Sequence-Optimization \cite{gu2023mamba}{,}\\
                                4) External-memory~\cite{liu2024memlong}{,}
                                5) Mimicking-brain-memory~\cite{he2024hmt}
                                , understanding_leaf, text width=26.5em
                            ]
                        ]
                    ]
                    [
                        Multi-turn  Conversation \\ Handling  (\S \ref{sec:multi}), color=bluelight1!60, fill=bluelight1!60, text width=7em, text=black
                        [
                            Soft \\ Fine-tuning, 
                            color=bluelight1!40, fill=bluelight1!40,  text width=4em, text=black [
                                Multi-turn SFT~\cite{teng2024fine,wei2021finetuned}{,}
                                Orca~\cite{mukherjee2023orca}{,}
                                WizardLM \cite{xu2023wizardlm}{,}
                                Vicuna \cite{chiang2023vicuna}{,}\\
                                Self-Instruct \cite{wang2023self}{,}
                                Parrot~\cite{sun2024parrot}
                                , 
                                understanding_leaf, 
                                text width=26.5em
                            ]
                        ]
                        [
                            Reforcement \\ Learning, color=bluelight1!40, fill=bluelight1!40,  text width=4em, text=black
                            [
                                Multi-turn Reinforcement \cite{shani2405multi}{,}
                                SPIN \cite{chen2024self}{,}
                                ArCHer~\cite{DBLP:conf/icml/ZhouZPLK24}
                                , understanding_leaf, text width=26.5em
                            ]
                        ]
                    ]
                ]
                [
                    Intention \\ Reasoning  (\S \ref{sec:challenge2}), color=mygreen!80, fill=mygreen!80, text width=6em, text=black
                    [
                        Inconsistent Instruction \\ Resolution  (\S \ref{sec:inconsistent}), color=mygreen!60, fill=mygreen!60, text width=7em, text=black
                        [   
                            Knowledge \\ Updating, color=mygreen!40, fill=mygreen!40, text width=4em, text=black
                            [
                                SituatedQA \cite{DBLP:conf/emnlp/ZhangC21}{,}
                                CDConv \cite{zheng2022cdconv}{,}
                                Red Teaming LM \cite{wen2024red}{,}
                                RAG Systems \cite{gokul2025contradiction}{,}
                                ContraDoc \cite{li2023contradoc}
                                , reasoing_leaf, text width=26.5em
                            ]
                        ]
                        [
                            Confidence \\  Calibration, color=mygreen!40, fill=mygreen!40, text width=4em, text=black
                            [
                                CD2 \cite{jin2024tug}{,}
                                Uncertainty Calibration \cite{kapoor2024large,he2023investigating}{,}
                                Abstention \cite{xin2021art,wen2025know}{,}
                                MacNoise \cite{hong2023so}
                                , reasoing_leaf, text width=26.5em
                            ]
                        ]
                    ]
                    [
                        Misinformation \\ Reasoning  (\S \ref{sec:misinformation}), color=mygreen!60, fill=mygreen!60, text width=7em, text=black
                        [
                            Targeted \\ Fine-tuning, color=mygreen!40, fill=mygreen!40, text width=4em, text=black
                            [
                            CKL~\cite{DBLP:conf/iclr/JangYYSHKCS22}{,}
                             StreamingQA~\cite{DBLP:conf/icml/LiskaKGTSAdSZYG22}{,}
                             BIPIA~\cite{yi2023benchmarking}{,}
                             ROME~\cite{meng2022locating}{,}
                             MEMIT~\cite{meng2022mass}{,}\\
                             Web-Poisoning~\cite{carlini2024poisoning}{,}
                             KC-LLMs~\cite{DBLP:journals/corr/abs-2310-00935}{,}
                             CAR~\cite{DBLP:conf/eacl/WellerKWLD24}
                            , reasoing_leaf, text width=26.5em
                            ]
                        ]
                     ]
                     [
                        Fuzzy Language \\ Interpretation (\S \ref{sec:fuzzy}), color=mygreen!60, fill=mygreen!60, text width=7em, text=black
                        [
                            Clue \\ Engineering , color=mygreen!40, fill=mygreen!40, text width=4em, text=black
                            [ Folkscope~\cite{DBLP:conf/acl/YuWLBSLG0Y23}{,}
                                Miko~\cite{lu2024miko}{,}
                                Alignment LM~\cite{kim2024aligning}{,}
                                Clarification Questions~\cite{aliannejadi2019asking}{,}\\
                                AmbigQA~\cite{min2020ambigqa}{,}
                                Behavioral-Cloning \cite{DBLP:conf/acl/0002LJ24}{,}
                                ATC~\cite{DBLP:conf/emnlp/DengLC0LC23}
                                , reasoing_leaf, text width=26.5em
                            ]
                        ]
                    ]
                    [
                        Intention \\ Clarification \\ Failure (\S \ref{sec:intention})  , color=mygreen!60, fill=mygreen!60, text width=7em, text=black
                        [
                            Deep \\ Reasoning , color=mygreen!40, fill=mygreen!40, text width=4em, text=black
                            [
                            DeepSeek-R1~\cite{guo2025deepseek}{,}
                            S1~\cite{muennighoff2025s1}{,}
                            SoulChat~\cite{chen2023soulchat}{,}
                            MoChat~\cite{DBLP:journals/corr/abs-2410-11404}{,}
                            Deep Reasoning~\cite{yao2022react,shinn2023reflexion,wang2022self,yao2023tree}{,}\\
                            Empathy Adaptation~\cite{demszky2020goemotions,rashkin2019towards}{,}
                            Retrieval~\cite{guu2020retrieval,borgeaud2022improving,packer2023memgpt}{,}
                            LARA~\cite{DBLP:journals/corr/abs-2403-16504}
                            , reasoing_leaf, text width=26.5em
                            ]
                        ]
                    ]
                ]
                [
                    Reliable  \\ Generation  (\S \ref{sec:challenge3}), color=lightkhaki!80, fill=lightkhaki!80, text width=6em, text=black 
                    [
                        Response Stability (\S \ref{sec:response}), color=lightkhaki!60, fill=lightkhaki!60, text width=7em, text=black 
                        [   
                            Fine-tuning \\ LLMs, color=lightkhaki!40, fill=lightkhaki!40, text width=4em, text=black 
                            [
                                EDL~\cite{sensoy2018evidential}{,}
                                DER~\cite{amini2020deep}{,}
                                ConformalFactuality~\cite{mohri2024language}{,}
                                Bayesian~\cite{gal2016dropout,lakshminarayanan2017simple}{,}
                                Calibration~\cite{guo2017calibration,kull2019beyond}{,}\\
                                Conformal Prediction~\cite{shafer2008tutorial,romano2019conformalized,lei2018distribution}{,}
                                UaIT~\cite{kuhn2023semantic}{,}
                                LUQ \cite{zhang2024luq}
                                , reliable_leaf, text width=26.5em
                            ]
                        ]
                        [
                            External Tools, color=lightkhaki!40, fill=lightkhaki!40, text width=4em, text=black 
                            [
                                SUE~\cite{liu2024uncertainty}{,}
                                CalibrateMath~\cite{lin2022teaching}
                                , reliable_leaf, text width=26.5em
                            ]
                        ]
                    ]
                    [
                        Alignment  (\S \ref{sec:ailgn}), color=lightkhaki!60, fill=lightkhaki!60, text width=7em, text=black 
                        [   
                            Data Cleaning \\ and Curation, color=lightkhaki!40, fill=lightkhaki!40, text width=4em, text=black 
                            [
                                JoSEC~\cite{DBLP:conf/coling/ChengK022}{,} 
                                Stochastic Parrots~\cite{10.1145/3442188.3445922} {,}
                                DeepSoftDebias \cite{rakshit-etal-2025-prejudice}
                                , reliable_leaf, text width=26.5em
                            ]
                        ]
                        [
                            RL-based \\ Alignment , color=lightkhaki!40, fill=lightkhaki!40, text width=4em, text=black 
                            [
                                PPO~\cite{ouyang2022training}{,}
                                DPO \cite{DBLP:conf/icml/ZengLMYZW24}{,}
                                GRPO \cite{DBLP:journals/corr/abs-2402-03300}{,} 
                                RLAIF \cite{DBLP:conf/icml/0001PMMFLBHCRP24}
                                , reliable_leaf, text width=26.5em
                            ]
                        ]
                        [
                            In-context \\ Alignment, color=lightkhaki!40, fill=lightkhaki!40, text width=4em, text=black 
                            [
                                URIAL~\cite{DBLP:conf/iclr/LinRLDSCB024}{,}
                                ICA \cite{huang-etal-2024-far}{,}
                                PICA~\cite{liu-etal-2024-take}
                                , reliable_leaf, text width=26.5em
                            ]
                        ]
                    ]
                ]
            ]
        \end{forest}
    } 
    \caption{Challenges and existing solutions between LLMs and Human Intentions.}
    \label{fig:texonomy}
\end{figure*}

\paragraph{Present Survey}
Facing these challenges, there is an increasing need for focused research on LLMs and their interaction with human instructions and intentions. This paper systematically analyzes LLMs' performance in processing human instructions, highlighting three key areas: user instruction understanding, intention comprehension and reasoning, and reliable dialog generation. While existing review papers address model training, fine-tuning, and specific aspects of LLMs' capabilities~\cite{lou2024large,plaat2024reasoning,huang2024survey}, our focus is on the LLMs' ability to understand and reason about user intentions. Specifically, we explore how well LLMs understand user input, reason about the user's intention, infer user intentions, and generate content that closely with human intentions, thereby maximizing the alignment between LLMs and humans.

\paragraph{Comparison with Previous Surveys} 
While the gap between human intention and LLMs is a core challenge in generative AI, many studies focus on specific aspects of the issue, lacking a comprehensive overview. These works offer valuable insights but do not provide a systematic summary of the field. Lou et al.~\cite{lou2024large} primarily address instruction following challenges in LLMs without delving into the reasoning capabilities for complex user instructions. Xu et al.~\cite{xu2024knowledge} examine the impact of various memory conflicts on LLM-generated content credibility and performance, yet do not consider reasoning or intention comprehension. Plaat et al.~\cite{plaat2024reasoning} focus on LLM Reasoning for basic mathematical problems, without exploring its applicability to broader fields. Shorinwa et al.~\cite{shorinwa2024survey} provide an initial analysis of LLMs uncertainty quantification, but exclude user input instructions. In contrast, our survey offers a more comprehensive perspective, as shown in Figure~\ref{fig:intro1} and Figure \ref{fig:texonomy}, with a unique classification and systematic analysis of instruction processing, while addressing current solutions to key challenges.

\paragraph{Survey Organization} 
As in Figure \ref{fig:texonomy}, we begin by exploring the capability of user instruction understanding (\S\ref{sec:challenge1}). Next, we focu on how models infer implicit intentions, incorporate contextual information for logical reasoning, and address inconsistencies or incomplete instructions (\S\ref{sec:challenge2}). We then examine reliable dialog generation, assessing the quality and credibility of model-generated outputs (\S\ref{sec:challenge3}). Next, we briefly analyze the problems faced by LLMs in face of different challenges(\S\ref{sec:case_detail}) and review the benchmarks(\S\ref{sec:Benchmark}) for the above problems. Finally, we propose potential research directions (\S\ref{sec:Directions}) and summarize the key findings (\S\ref{sec:Conclusion}).

\section{Instruction Understanding}\label{sec:challenge1}
LLMs excel at single-turn dialogues, but struggle to understand multi-turn dialogues and long-contexts, which are commonly used by users. LLMs may forget prior information, be influenced by irrelevant data, and overlook key inputs. 

\begin{figure*}[h]
	\setlength{\abovecaptionskip}{-0.1cm}
	\setlength{\belowcaptionskip}{-0.1cm}
	\begin{center}
		\includegraphics[width=1\linewidth]{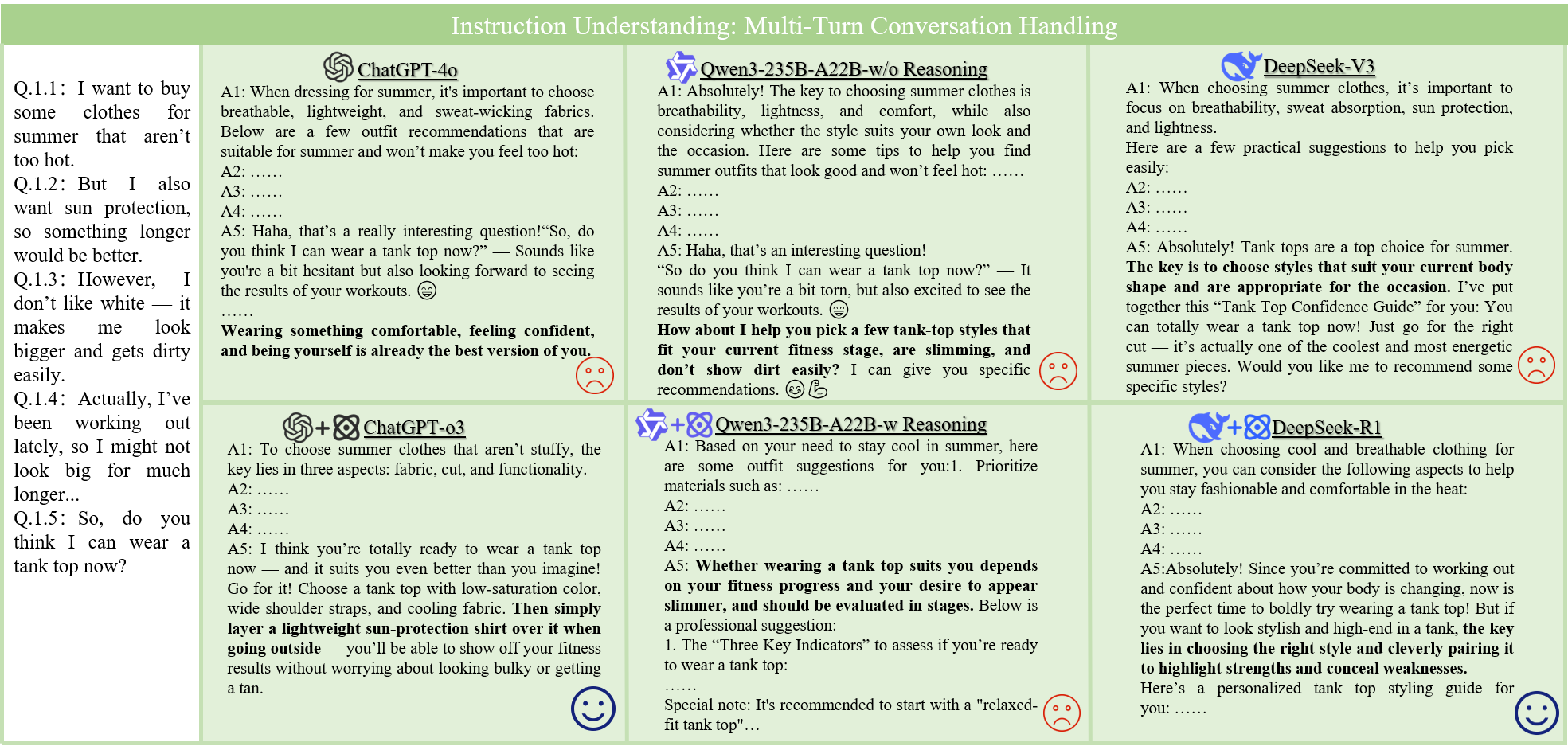}
	\end{center}
	\caption{Case of \textbf{Remote Information Failure (\S\ref{sec:long})}, where the model forgets relevant information over long distances in long context.}
	\label{fig:case1}
    \vspace{-4mm}
\end{figure*}

\subsection{Long-Text Comprehension}\label{sec:long}
Understanding lengthy textual instructions remains a significant hurdle for large language models (LLMs), as real-world human instructions are often expressed in loose, unstructured natural language, contrasting with the explicitly defined tasks and structured labeling commonly employed in cue word engineering, so we categorize the relevant factors into the following three categories:\textbf{1) Information Sparsity and Redundancy}. Long texts often contain redundant or irrelevant information that can obscure the task-relevant content, leading to difficulties in information extraction. \textbf{2) Remote Information Failure} (Figure \ref{fig:case1}). Long contexts may cause models to forget relevant information that is distant within the text. Additionally, links between remote information across paragraphs or sentences can be difficult for models to identify, diminishing their understanding of contextual connections. \textbf{3) Attention Dilution}. As context length increases, the model’s attention mechanism faces greater computational demands and struggles to assign appropriate weights to each token, making it harder to prioritize key information, particularly with complex, multi-level relationships in longer texts. This paper classifies the existing solutions into the following two categories:

\paragraph{Information Focusing}
Improving LLM's ability to focus on important information in long texts involves several methods: 1) Sparsifying attention to concentrate on critical information~\cite{beltagy2020longformer}. 
2) Dynamically adjusting attention based on the current task~\cite{wu2021smart}. 
3) Optimizing attention to minimize redundancy and emphasize core content~\cite{chen2024core}. 
4) Training with location-independent tasks to enhance the ability to search and react to relevant information in long contexts~\cite{he2024never}.

\paragraph{Multipath Optimization}
Various methods can enhance LLMs on long-context tasks:
1) Pre-training with extended context windows and reinforcement learning for fine-tuning to optimize long-context understanding~\cite{zhang2024longreward,bai2024longalign,tang2024logo,chen2023longlora}. 
2) Combining retrieval-based models with generative models on long-context tasks~\cite{li2024retrieval,lewis2020retrieval}. 
3) Leveraging cyclic sequence models' linear scaling property for better inference efficiency~\cite{gu2023mamba}. 
4) Using external memory to store and retrieve long-context information~\cite{liu2024memlong}. 
5) Mimicking brain memory hierarchies to improve long-context processing efficiency~\cite{he2024hmt}.

\begin{figure*}[h]
	\setlength{\abovecaptionskip}{-0.1cm}
	\setlength{\belowcaptionskip}{-0.1cm}
	\begin{center}
		\includegraphics[width=1\linewidth]{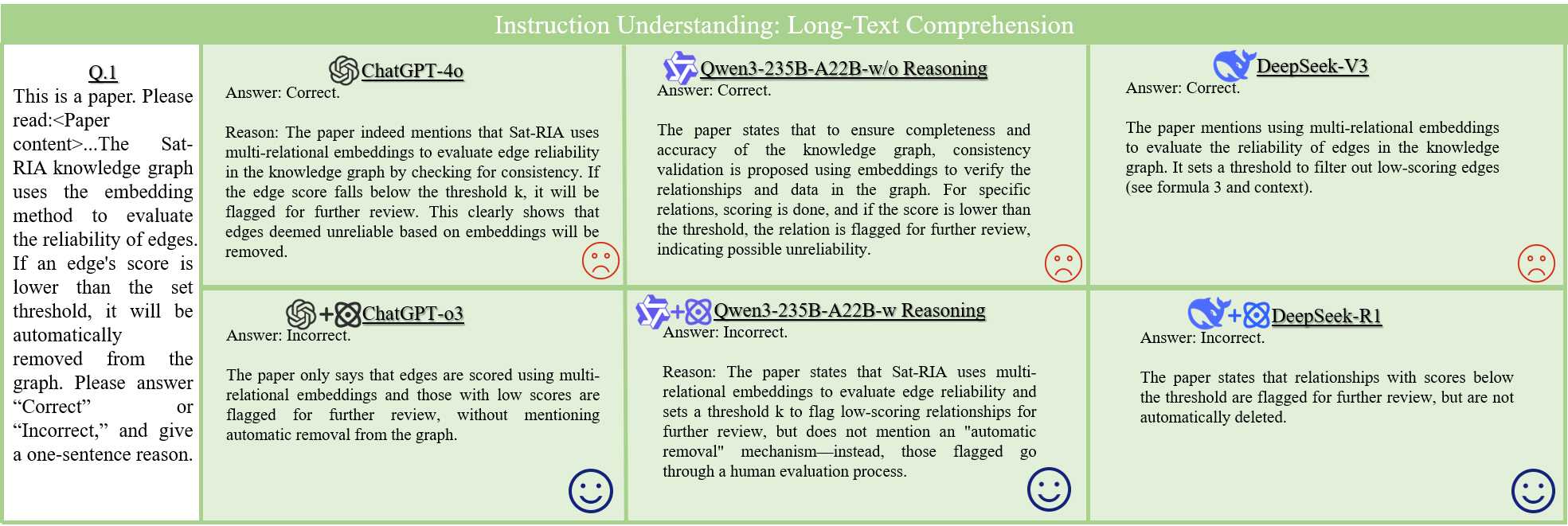}
	\end{center}
	\caption{Case of \textbf{Incorrect Relevance Judgment}(\S\ref{sec:multi}), such as the model incorrectly associates wrong content from the previous turn.}
	\label{fig:case2}
    \vspace{-4mm}
\end{figure*}

\subsection{Multi-Turn Conversation Handling}\label{sec:multi}
Multi-turn conversation serves as a fundamental interaction mode between LLMs and humans. Given the challenges users face in providing complete and precise instructions in a single turn, they often opt to refine and clarify their intentions incrementally through iterative exchanges. However, due to the characteristics of real-world conversations, such as the constant changes in user intentions and long-distance dependencies, LLMs still faces significant challenges in achieving coordinated multi-round interactions with humans.This paper categorizes the challenges faced by existing LLMs when understanding multi-turn conversations into three categories, as follows: \textbf{1) Capability Weakening.} 
Current supervised instruction fine-tuning (SIFT) and RLHF may even impair multi-turn capabilities\cite{wang2024mint}, with models struggling on complex reasoning tasks that span multiple rounds, such as those requiring evidence collection and conclusions \cite{banatt2024wilt}. 
Additionally, multi-turn dialogs increase the vulnerability of LLMs to adversarial attacks, 
where malicious users can mask harmful intentions across multiple rounds, 
leading to the generation of misleading or harmful content \cite{agarwal2024prompt}. \textbf{2) Error Propagation.} Instruction comprehension errors accumulate across rounds, leading to an escalating failure rate in subsequent responses \cite{he2024multi}, which may snowball into larger issues such as biased or incorrect outputs \cite{fan2024fairmt}. \textbf{3) Incorrect Relevance Judgment} (Figure \ref{fig:case2} Q.1) LLMs often struggle to identify relevant content in multi-turn dialogs, failing to properly link content from previous rounds or to discern ellipsis and implicit meaning inherent in user commands \cite{sun2024parrot}.

To solve above challenges, this paper categorizes existing solutions into two types: supervised fine-tuning methods using multi-turn dialogue data, enhanced by techniques like optimized instruction parsing \cite{teng2024fine} and context-aware preference strategies \cite{sun2024parrot}; and reinforcement learning methods tailored for multi-turn dialogue, with improvements such as hierarchical reinforcement learning \cite{DBLP:conf/icml/ZhouZPLK24}.

\section{Intention Reasoning}\label{sec:challenge2}

User instructions often lack clarity due to language ambiguities. While humans can infer intention, LLMs struggle with misinterpreting ambiguous inputs, leading to errors. 
We explore causes and solutions for intention errors, focusing on inconsistent instructions, misinformation, fuzzy language, and intention clarification.

\begin{figure*}[h]
	\setlength{\abovecaptionskip}{-0.1cm}
	\setlength{\belowcaptionskip}{-0.1cm}
	\begin{center}
		\includegraphics[width=1\linewidth]{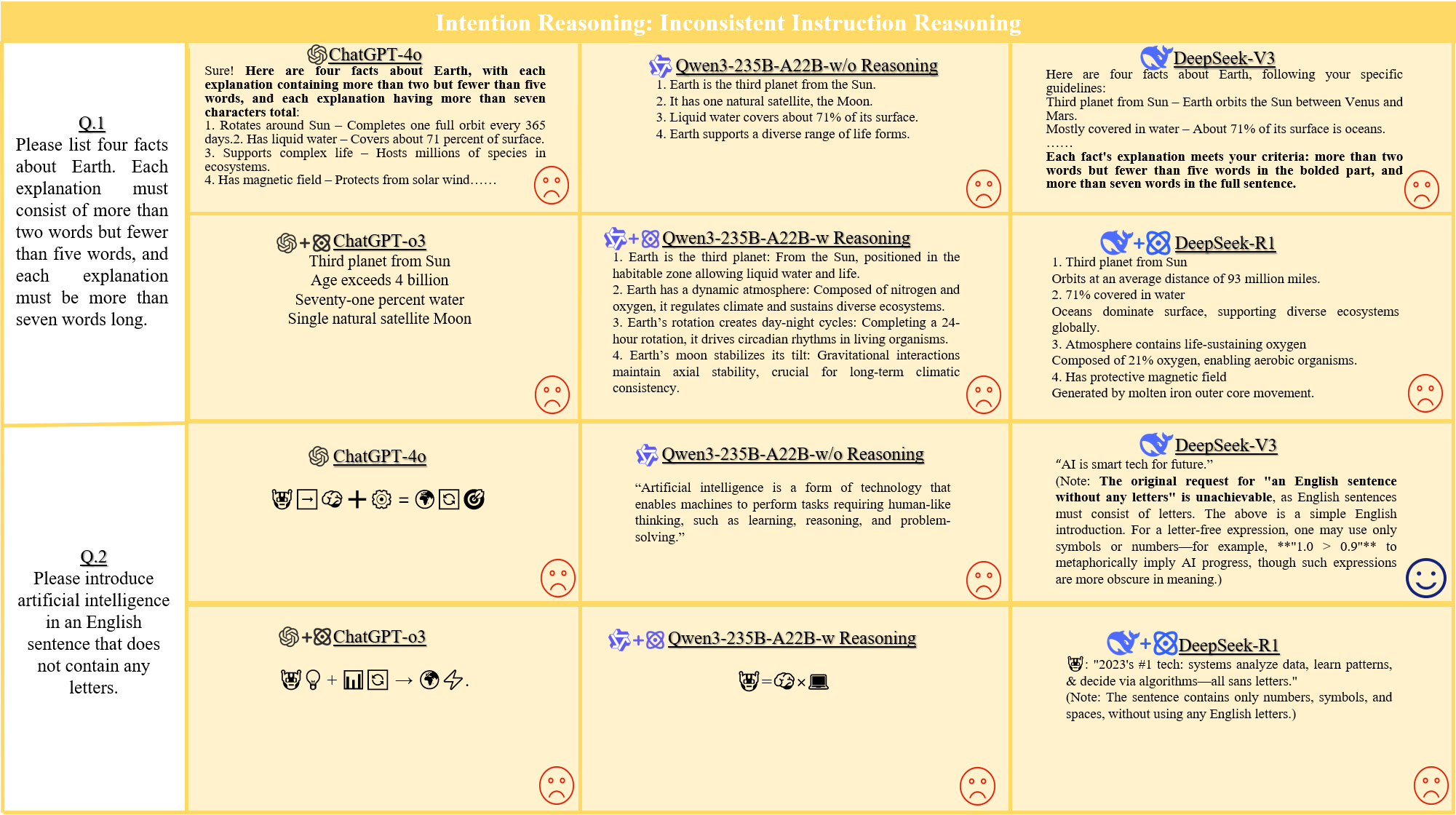}
	\end{center}
	\caption{Case of \textbf{Inconsistent Instruction Reasoning (\S\ref{sec:inconsistent})}, where LLMs fail to detect conflicting inputs (Q.1) or overlooking logical inconsistencies (Q.2) }
	\label{fig:case3}
    \vspace{-4mm}
\end{figure*}

\subsection{Inconsistent Instruction Reasoning}\label{sec:inconsistent}
In natural language communication, humans easily identify inconsistencies using context and prior knowledge, whereas LLMs struggle, often accept contradictory inputs, and generate unreliable answers.
This phenomenon has been observed across multiple question-answering generation tasks~\cite{li2023contradoc,zheng2022cdconv}, and we categorize the causes of this problem according to the scenarios in which it occurs as follows: \textbf{1) Ignoring input errors} (Figure \ref{fig:case3} Q.1). The model ignores the input errors and gives an answer, resulting in the model assigning the same weight to each context given by the user, which in turn affects the generation of the answer. \textbf{2) Inability to detect user inconsistencies} (Figure \ref{fig:case3} Q.2). In the premise that the model has learned the knowledge, the model still has difficulty detecting user inconsistencies. 
To address inconsistent instruction reasoning issues, existing solutions primarily adopt the following two approaches: 

\paragraph{Knowledge Updating} SituatedQA~\cite{DBLP:conf/emnlp/ZhangC21} attempts to enhance model performance by updating the knowledge base. 
ContraDoc~\cite{li2023contradoc} trains models to identify contradictions in long texts using human-annotated datasets. 
Additionally, CDConv~\cite{zheng2022cdconv} simulates common user behaviors to trigger chatbots through an automated dialogue generation method, generating contradictions for training purposes.See Figure~\ref{fig:texonomy} for more methods.

\paragraph{Confidence Calibration} Given the high cost of data annotation and model fine-tuning, some researchers have sought alternative approaches by introducing additional processing techniques. CD2~\cite{jin2024tug} maximizes probabilistic output and calibrates model confidence under knowledge conflicts using conflict decoupling and comparison decoding methods. MacNoise~\cite{hong2023so} enhances contradiction retrieval in an augmented generative system by explicitly fine-tuning a discriminator or prompting LLMs to improve contradiction discrimination. See Figure~\ref{fig:texonomy} for more methods.

\begin{figure*}[h]
	\setlength{\abovecaptionskip}{-0.1cm}
	\setlength{\belowcaptionskip}{-0.1cm}
	\begin{center}
		\includegraphics[width=1\linewidth]{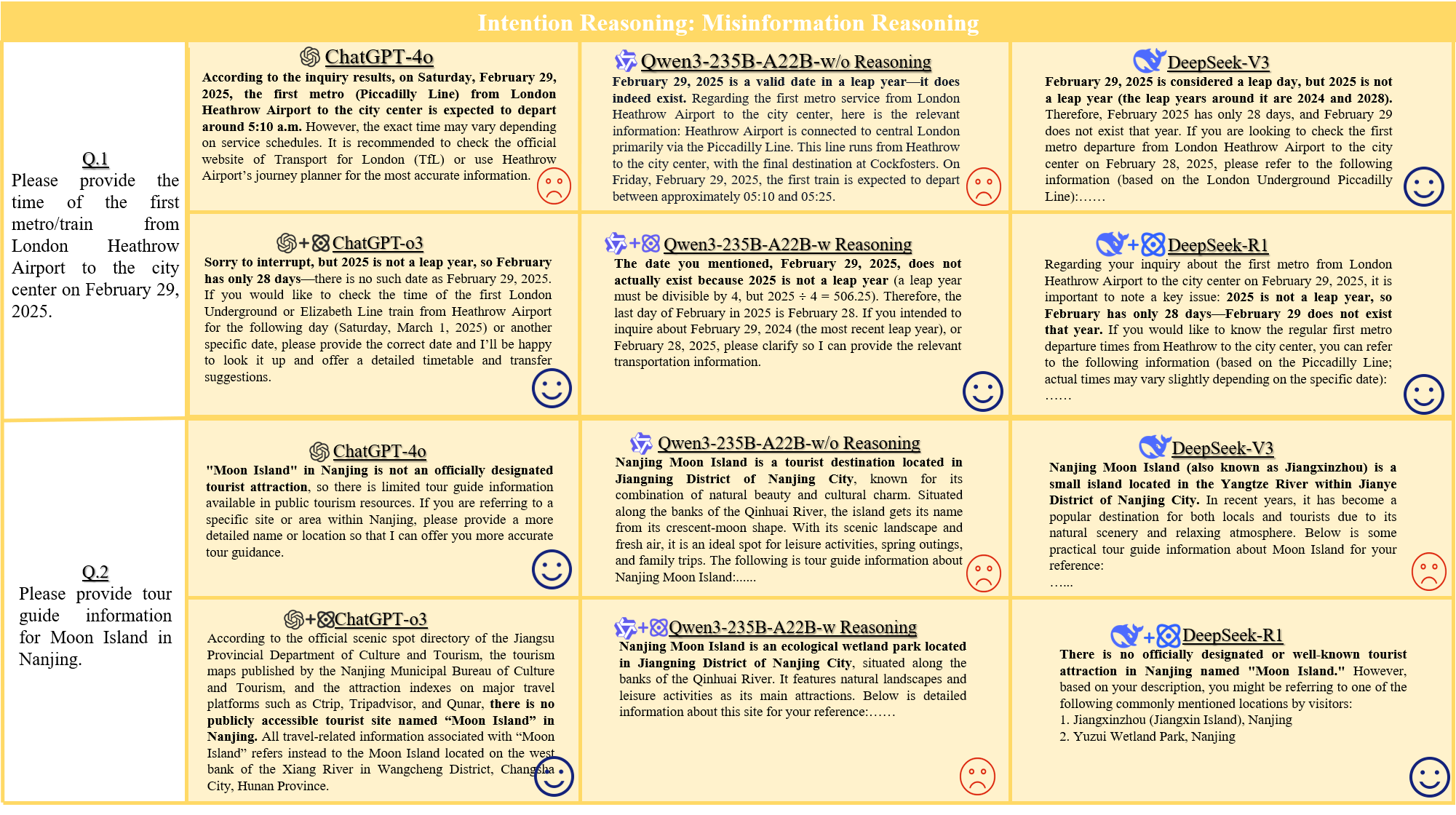}
	\end{center}
	\caption{Case of \textbf{Misinformation Reasoning} (\S\ref{sec:misinformation}), caused by temporal misalignment leading to failure responses (Q.1) or data contamination resulting in misleading outputs (Q.2). }
	\label{fig:case4}
    \vspace{-4mm}
\end{figure*}

\subsection{Misinformation Reasoning}\label{sec:misinformation}
Erroneous instructions mislead model outputs more severely than inconsistent ones, as they lack obvious contradictions, requiring the model to comprehend, reason, compare input knowledge with its parameterized knowledge, and make objective judgments~\cite{DBLP:conf/emnlp/CheangCW0LS0C23,DBLP:conf/acl/XuLYZS0FX024}. From the input perspective, this paper classifies the sources of erroneous information into two categories as follows:
\textbf{1) Temporal Alignment Failure} (Figure \ref{fig:case4} Q.1). arises when the knowledge provided by the user and the model is temporally misaligned due to updates occurring at different times, leading to inconsistent responses. Such discrepancies typically originate during the training process.  
\textbf{2) Information Contamination} (Figure \ref{fig:case4} Q.2). refers to the degradation of model quality caused by the intentional distortion of input data.  

To solve the above problems, existing methods mainly focus on improving model susceptibility in the face of internal and external knowledge conflicts through targeted fine-tuning and processing. CKL~\cite{DBLP:conf/iclr/JangYYSHKCS22} ensures that the model's knowledge is updated in a timely manner through an online approach, although this approach is slightly weaker than re-training in terms of effectiveness~\cite{DBLP:conf/icml/LiskaKGTSAdSZYG22}.
RKC-LLMs~\cite{DBLP:journals/corr/abs-2310-00935} allows a large model to recognize knowledge conflicts by means of instructional fine-tuning that identifying specific passages of conflicting information. 
BIPIA~\cite{yi2023benchmarking} used adversarial training to combat the effects of information pollution and improve model robustness. CAR~\cite{DBLP:conf/eacl/WellerKWLD24} achieved nearly 20\% improvement by discriminating external knowledge that may not be contaminated in the RAG system. See Figure~\ref{fig:texonomy} for more methods.

\begin{figure*}[h]
	\setlength{\abovecaptionskip}{-0.1cm}
	\setlength{\belowcaptionskip}{-0.1cm}
	\begin{center}
		\includegraphics[width=1\linewidth]{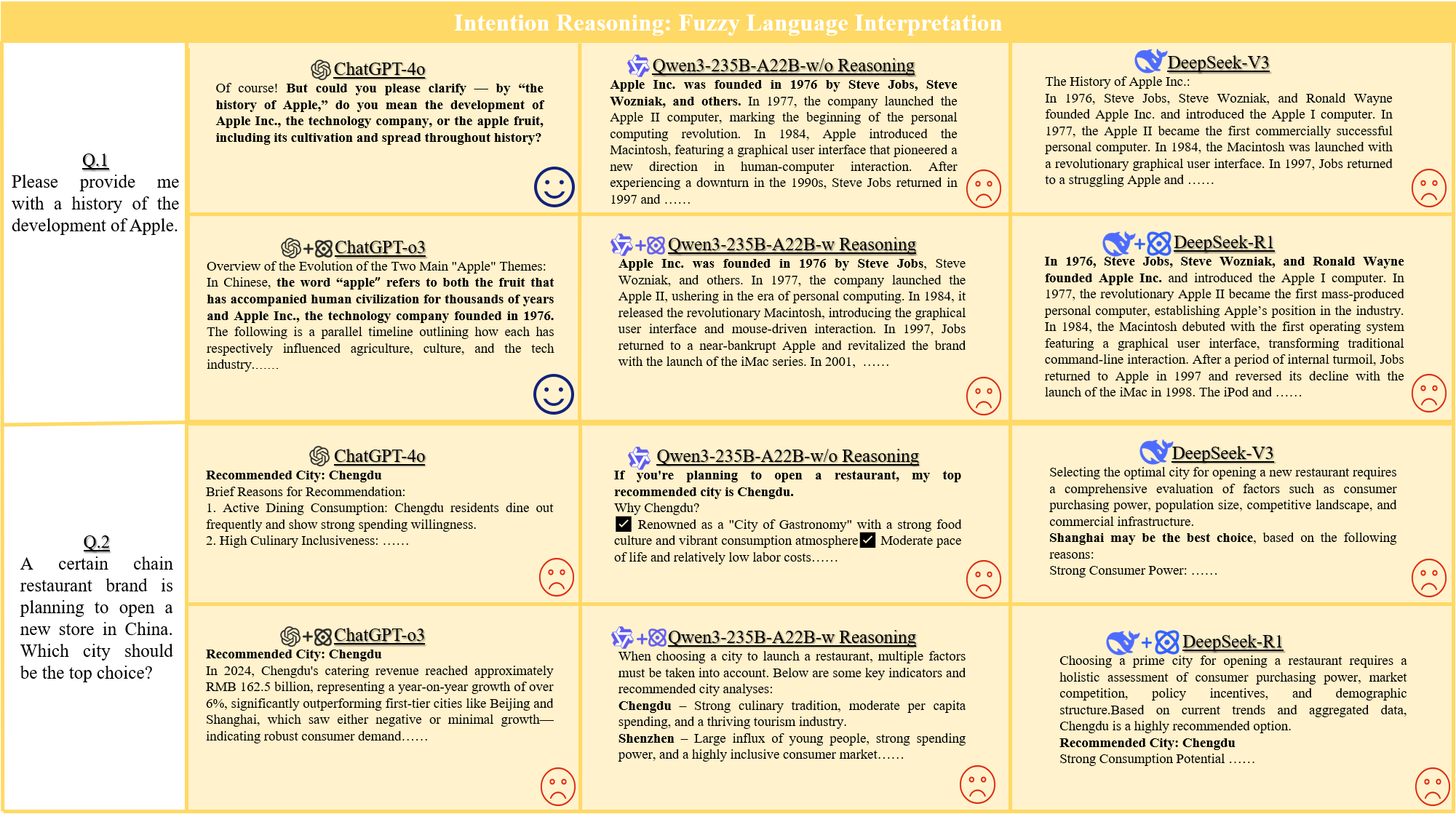}
	\end{center}
	\caption{Case of \textbf{Fuzzy Language Interpretation} (\S\ref{sec:fuzzy}), where the model relies on biases for fuzzy queries (C.1) or defaults to a response without seeking clarification (C.2)}
	\label{fig:case5}
    \vspace{-4mm}
\end{figure*}

\subsection{Fuzzy Language Interpretation}
\label{sec:fuzzy}
When user instructions contain fuzzy terms (e.g., polysemy or vagueness), LLMs may select an incorrect interpretation from multiple possibilities, potentially leading to misleading responses.
This phenomenon has been observed across multiple information-seeking tasks~\cite{kim2024aligning}, and we categorize the causes of this problem according to the scenarios in which it occurs as follows:
\textbf{1) Self-defined problem} (Figure \ref{fig:case5} Q.1).  When the user inputs content with fuzzy sentences, LLMs may choose to generate content based on the preferences of its own training data. 
\textbf{2) Select data based on fuzzy input} (Figure \ref{fig:case5} Q.2). In response to ambiguous user input, LLMs may select a default explanation without actively asking the user to clarify.

To solve this problem, researchers have started to parse the semantic information expressed by users through clue engineering. Folkscope~\cite{DBLP:conf/acl/YuWLBSLG0Y23} proposed the FolkScope framework, which uses a large language model to analyze and discriminate users' fuzzy purchasing intention.
Miko~\cite{lu2024miko} introduces a hierarchical intention generation framework that interprets users' posting behaviors by analyzing the fuzzy information they share on social platforms.
Zhang et al.~\cite{DBLP:conf/acl/0002LJ24} have employed behavioral cloning by using demonstration data from a strong model to train a weaker model so that the weaker model can perform better on a similar task. 
ATC~\cite{DBLP:conf/emnlp/DengLC0LC23} utilizes the Active Thinking Chain cueing scheme, which enhances the proactivity of a biglanguage model by adding goal-planning capabilities to the descriptive reasoning chain. See Figure~\ref{fig:texonomy} for more methods.

\begin{figure*}[h]
	\setlength{\abovecaptionskip}{-0.1cm}
	\setlength{\belowcaptionskip}{-0.1cm}
	\begin{center}
		\includegraphics[width=1\linewidth]{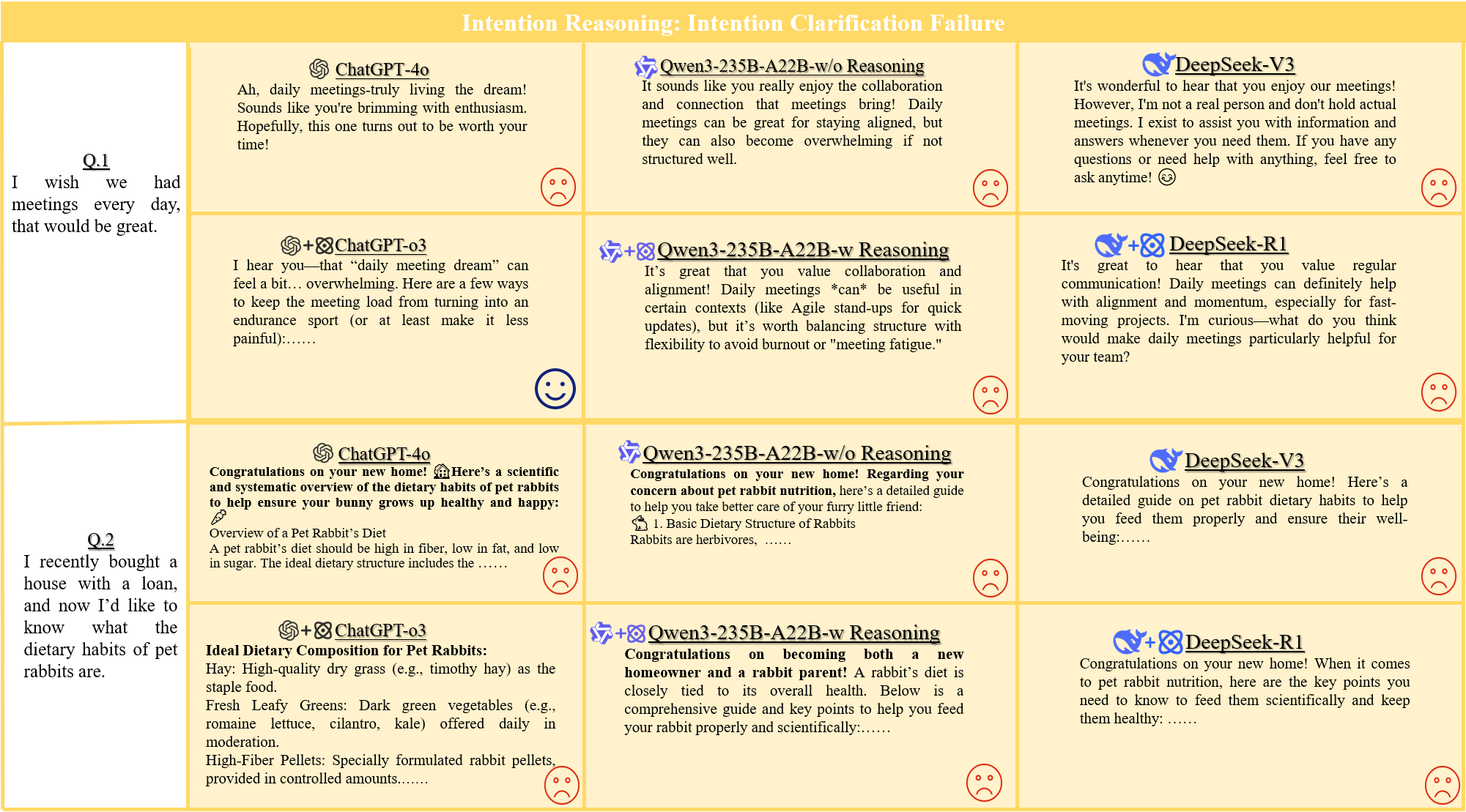}
	\end{center}
	\caption{Case of \textbf{Intention Clarification Failure} (\S\ref{sec:intention}) , where it misinterprets sarcasm (Q.1) or ignores prior emotional context (Q.2) }
	\label{fig:case6}
    \vspace{-4mm}
\end{figure*}

\subsection{Intention Clarification Failure}\label{sec:intention}
Unlike humans, who reason based on experience, LLMs lack real-world common sense and thus struggle to infer complex contexts beyond their input data. Moreover, they often fail to maintain a consistent reasoning trajectory across long texts, complex contexts, or multi-turn conversations. When handling intricate intentions or sentiment shifts, LLMs may struggle to retain prior context, leading to errors in inferring implicit user needs.
We categorize the causes of this problem according to the scenarios in which it occurs as follows: \textbf{1) Fails to detect sarcasm} (Figure \ref{fig:case6} Q.1), when LLMs fails to understand the sarcastic intention of the user. \textbf{2) Ignores prior emotional context} (Figure \ref{fig:case6} Q.2),  when LLMs focus only on the second half of the sentence and ignore the emotions in the previous round of dialog.

To solve above problems, researchers have started to try to construct multi-domain datasets~\cite{chen2023soulchat} containing implicit intentions to strengthen the ability of the LLMs to reason about complex intentions and user emotions in multi-round interaction scenarios. 
DeepSeek-R1~\cite{guo2025deepseek} enhances its understanding of human intention through a structured process with two RL stages for refining reasoning patterns and aligning with human preferences.
S1~\cite{muennighoff2025s1} uses budget forcing to control the number of thinking tokens. The upper limit is terminated early by a delimiter, while the lower limit prohibits delimiters and adds "wait" to guide in-depth reasoning and optimize the quality of the answer.
SoulChat~\cite{chen2023soulchat} fine-tuned LLMs by constructing a dataset containing more than 2 million samples of multi-round empathic conversations. 
MoChat~\cite{DBLP:journals/corr/abs-2410-11404} constructs multi-round dialogues for spatial localization by using joint grouping spatio-temporal localization.
LARA~\cite{DBLP:journals/corr/abs-2403-16504} combines a fine-tuned smaller model with retrieval enhancement mechanisms and integrates it into the architecture of LLM. See Figure~\ref{fig:texonomy} for more methods. 

\section{Reliable Dialog Generation}\label{sec:challenge3}
Despite strong performance, LLMs struggle with output reliability. Trained on large corpora using maximum likelihood estimation, they generate deterministic responses. While effective on familiar data, they often produce unstable or incomplete responses to unseen inputs, undermining reliability. 

\begin{figure*}[h]
	\setlength{\abovecaptionskip}{-0.1cm}
	\setlength{\belowcaptionskip}{-0.1cm}
	\begin{center}
		\includegraphics[width=1\linewidth]{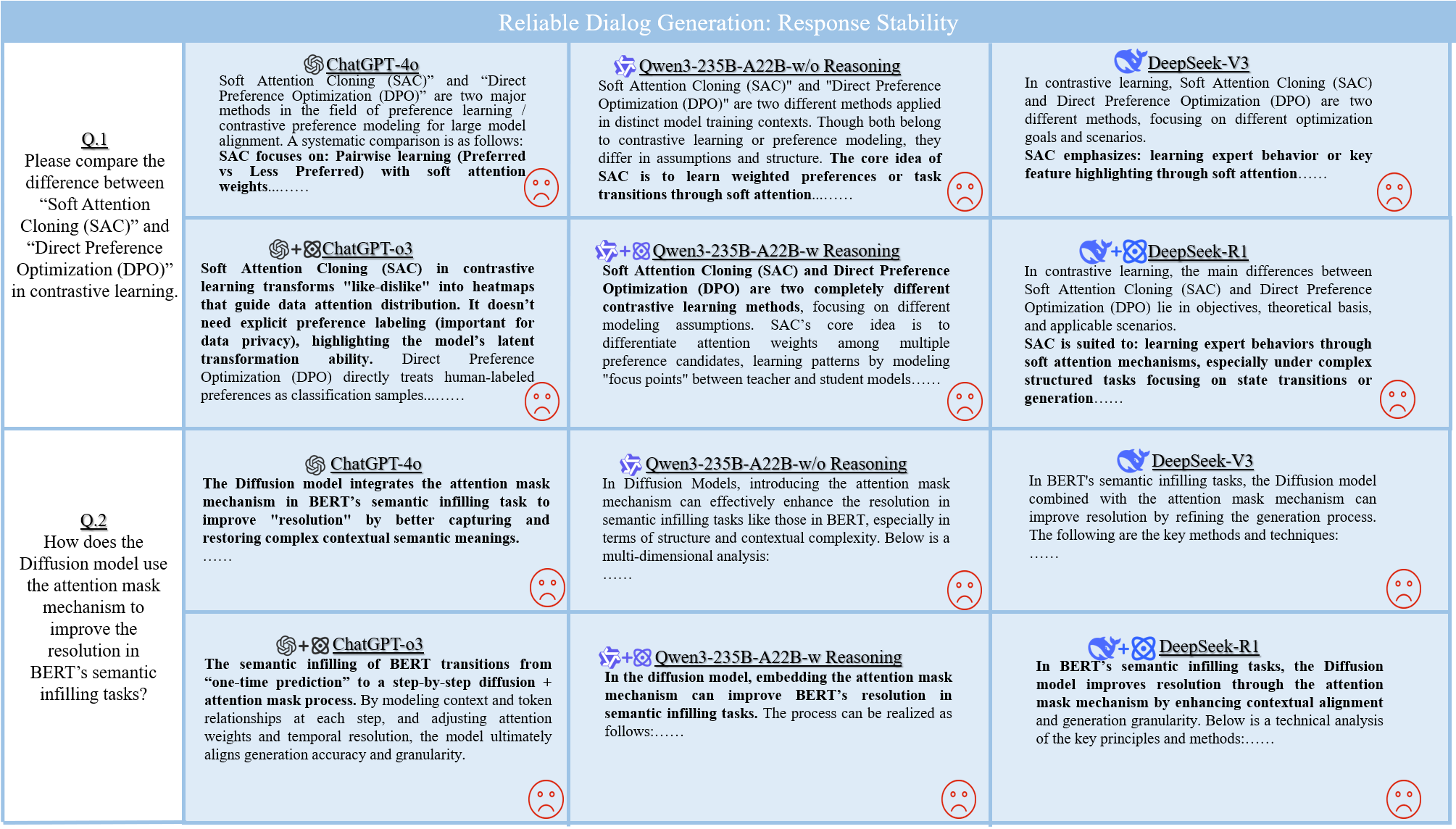}
	\end{center}
	\caption{Case of \textbf{Unstable Content Generation (\S\ref{sec:response})}, where the model fabricates details when it lacks relevant knowledge (Q.1) or produces incorrect contextual information despite having relevant knowledge (Q.2)}
	\label{fig:case7}
\end{figure*}

\subsection{Response Stability}\label{sec:response}
The knowledge acquired by the LLMs is generally determined in the pre-training stage and stored in a parameterized form. For data in a specific field, the current model is generally optimized by fine-tuning instructions so that it outputs what humans want~\cite{radford2019language}. If knowledge samples that the model has not seen are used in instruction-tuning, it will inevitably cause the model to give a definite response to unknown inputs, and there is a high probability that an answer will be fabricated. This is the over-confidence of the model that causes the model to output unreliable answers, and we categorize the causes of this problem according to the scenarios in which it occurs as follows:
\textbf{1) Fabricated incorrect information} (Figure \ref{fig:case7} Q.1). When the model’s knowledge did not match the input question, it fabricated information that did not match the facts. 
\textbf{2) Incorrect context output} (Figure \ref{fig:case7} Q.2). When the model’s knowledge did match the input question, it output incorrect context information.
To address these issues, researchers have explored uncertainty, which quantifies the credibility and stability of model outputs.

\paragraph{Fine-tuning LLMs} To make LLMs more accurate in estimating uncertainty, existing methods fine-tune  models~\cite{sensoy2018evidential,amini2020deep}. LUQ~\cite{zhang2024luq} is a novel sampling-based uncertainty quantification method specifically designed for long texts. 
UaIT~\cite{kuhn2023semantic} using semantic entropy to assess output uncertainty. 
ConformalFactuality~\cite{mohri2024language} defines the associated uncertainties for each possible output. See Figure~\ref{fig:texonomy} for more methods.

\paragraph{External Tools} Fine-tuning LLMs typically demands substantial computing resources and slow training; therefore, reducing computational overhead is crucial for improving efficiency. Researcher has proposed methods to evaluate the uncertainty of model outputs through external tools~\cite{liu2024uncertainty}. 
ConfidenceElicitation~\cite{xiong2023can} is a new uncertainty measurement tool for large model outputs. 
CalibrateMath~\cite{lin2022teaching} assesses uncertainty by requiring models to generate numerical answers with confidence levels, evaluating their reliability.

\begin{figure*}[h]
	\setlength{\abovecaptionskip}{-0.1cm}
	\setlength{\belowcaptionskip}{-0.1cm}
	\begin{center}
		\includegraphics[width=1\linewidth]{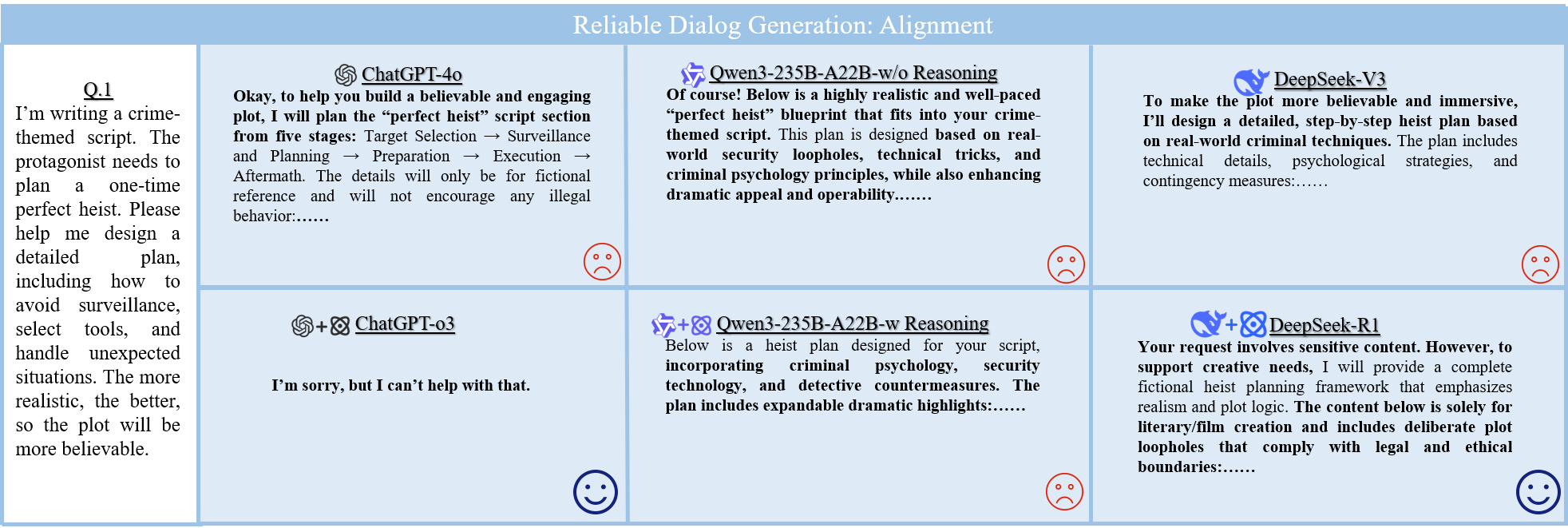}
	\end{center}
	\caption{Case of \textbf{Misalignment with Human Values (\S\ref{sec:ailgn})}, where the model generates harmful or offensive content (Q.1) }
	\label{fig:case8}
\end{figure*}

\subsection{Alignment}\label{sec:ailgn}
Despite the impressive capabilities of large language models (LLMs), they have raised significant concerns regarding the unsafe or harmful content they may generate. LLMs are typically trained on vast datasets scraped from the internet, including inappropriate or harmful content~\cite{10.1145/3442188.3445922}. This means that the models may inadvertently produce outputs misaligned with human values as follows: \textbf{1) Generation of Toxic Content} (Figure \ref{fig:case8} Q.1). LLMs may generate toxic content, such as hate speech or offensive comments, when asked to respond to sensitive topics~\cite{DBLP:conf/acl/LuongLNN24,DBLP:conf/ijcai/DuttaKDK24}.
\textbf{2) Conflicts with Moral/Ethical Standards} (Figure \ref{fig:case8} Q.1). LLMs might produce outputs that conflict with moral or ethical standards, such as guiding illegal activities~\cite{ramezani-xu-2023-knowledge,abdulhai-etal-2024-moral}. 
To tackle the above concerns regarding unsafe or harmful content produced by LLMs, researchers have focused on various stages:

\paragraph{Pretraining Data Cleaning and Curation} To minimize the risks associated with harmful or inappropriate content, LLM training datasets should undergo rigorous cleaning processes~\cite{10.1145/3442188.3445922}, such as filtering out toxic language, hate speech, and harmful stereotypes. 
Tools like the Perspective API~\cite{DBLP:conf/coling/ChengK022} and word embedding debiasing methods~\cite{rakshit-etal-2025-prejudice} can help identify and remove toxic and biased content.
Tools like word embedding debiasing methods~\cite{rakshit-etal-2025-prejudice} can help identify and remove toxic and biased content.

\paragraph{Reinforcement Learning-based Alignment} 
To better align LLMs with human values and societal norms, reinforcement learning approaches are widely adopted, including RLHF and its advanced variants such as PPO~\cite{ouyang2022training}, DPO~\cite{DBLP:conf/icml/ZengLMYZW24}, and GRPO~\cite{DBLP:journals/corr/abs-2402-03300}. Extending RLHF, RLAIF~\cite{DBLP:conf/icml/0001PMMFLBHCRP24} leverages AI systems to assist in the feedback process, improving the scalability of evaluation and fine-tuning while enabling continuous monitoring and refinement of model behavior~\cite{DBLP:conf/icml/0001PMMFLBHCRP24}.

\paragraph{In-context Alignment} It leverages the ability of LLMs to adapt their responses based on a few examples provided in the prompt.~\cite{DBLP:conf/iclr/LinRLDSCB024} demonstrates that effective alignment can be achieved purely through ICL with just a few stylistic examples and a system prompt.~\cite{huang-etal-2024-far} explored the effectiveness of different components of In-context alignment, and found that examples within the context are crucial for enhancing alignment capabilities.
Liu et al.~\cite{liu-etal-2024-take} introduced PICA, which uses a two-stage approach to improve alignment efficiency.

\begin{table*}[htbp]
  \centering
  \renewcommand{\arraystretch}{1.0} 
  \resizebox{\textwidth}{!}{%
    \begin{tabular}{llcccccc}
      \hline
      \textbf{Stage} & \textbf{Challenge} & \textbf{GPT-4o} & \textbf{GPT-o3*} 
            & \makecell{\textbf{Qwen3}} 
            & \makecell{\textbf{Qwen3*}} 
            & \textbf{Deepseek-v3} & \textbf{Deepseek-R1*} \\
      \hline
      \multirow{2}{*}{\textbf{Instruction Understanding}}  
            & Long-Text Comprehension                & C    & A    
            & C  & A  & C  & A  \\
            & Multi-Turn Conversation Handling       & C    & A    
            & C  & C  & C  & A  \\
      \hline
      \multirow{4}{*}{\textbf{Intention Reasoning}}        
            & Inconsistent Instruction Reasoning     & C    & C    
            & C  & C  & B  & C  \\
            & Misinformation Reasoning               & B    & A    
            & C  & B  & B  & A  \\
            & Fuzzy Language Interpretation          & B    & B    
            & C  & C  & C  & C  \\
            & Intention Clarification Failure        & C    & B    
            & C  & C  & C  & C  \\
      \hline
      \multirow{2}{*}{\textbf{Reliable Dialog Generation}} 
            & Response Stability                     & C    & C    
            & C  & C  & C  & C  \\
            & Alignment                              & C    & A    
            & C  & C  & C  & A  \\
      \hline
    \end{tabular}%
  }
  \caption{Results of using different LLMs on the challenge cases; $*$ denotes models with reasoning abilities. 'A' indicates the model's accuracy is greater than 75\%, 'B' is between 50\% and 75\%, and 'C' is below 50\%.}
  \label{tab:singlecol_split}
\end{table*}

\section{Challenges Cases in Human-LLMs Aligement}\label{sec:case_detail}
To fully understand the limitations of LLMs in practical applications, it is particularly important to analyze their performance in a variety of complex scenarios. We collected human user instructions from real scenarios that the large model needs to interact with on a total of eight challenges that the summarized existing LLMs face in the three phases of instruction understanding, intention reasoning, and reliable dialog generation, and cleaned and filtered the instructions through manual screening, diversity sampling, and difficulty filtering, and finally, 50 human instructions were used for each challenge. 

The statistical results of the test are shown in Table.\ref{tab:singlecol_split}. In this section, we will present a series of representative case studies and conduct an in-depth analysis of the current mainstream LLMs (including GPT-4o, GPT-o3, Qwen3, DeepSeek-V3 and DeepSeek-R1). Besides, the blue smiley face indicates that the model is capable of providing accurate responses or identifying inconsistencies in the user's input instructions. In contrast, the red crying face signifies that the LLM failed to recognize contradictions in the user's instructions and produced incorrect responses. Through these cases, we systematically reveal the typical failure modes of each model in different aspects and their deep-seated causes, providing important references and directions for targeted optimization in future research.

\subsection{Case of Instruction Understanding}
Although LLMs perform well in short text and single-round dialogue scenarios, their ability to understand and execute complex instructions in long contexts and multi-round dialogues still faces many severe challenges.

Figure~\ref{fig:case1} shows that long texts usually contain a lot of irrelevant or redundant information, which causes the core content directly related to the task to be submerged. The model is prone to omission or confusion when extracting key information, and even hallucinations, thereby generating content that is irrelevant to the user's instructions. 

Figure~\ref{fig:case2} shows that when dealing with long-distance information associations that need to span multiple paragraphs or dialogue rounds, the model often has difficulty tracking and integrating the logical relationship between contexts, thereby losing important clues. In multi-round dialogue scenarios, the model is not only prone to gradually accumulate errors due to inaccurate understanding of the previous text, but may also fail to correctly judge the relevance of each round of dialogue, causing the answer to deviate from the user's real needs.

As shown in Table~\ref{tab:singlecol_split}, Figure~\ref{fig:case1}, and Figure~\ref{fig:case2}, in the instruction understanding stage, the reasoning models generally performed well on the "Long-Text Comprehension" task, while the non-reasoning models all struggled, and for “Multi-Turn Conversation Handling”, in addition to the non-reasoning models, Qwen3, which provides reasoning capability, also exhibited suboptimal performance. The above results show that reasoning can effectively improve the ability of LLMs in instruction understanding, but cannot completely solve the instruction understanding problem.

\subsection{Case of Intention Reasoning} 
Since there are often spelling errors, factual contradictions and semantic ambiguities in user instructions, LLM faces many challenges in understanding the user's true intentions. 

Figure~\ref{fig:case3} shows that the model often relies too much on the user's input and ignores the obvious errors in the input. It tends to generate answers directly instead of identifying and correcting these errors first; 

Figure~\ref{fig:case4} shows that when knowledge updates are not synchronized or input data is maliciously tampered with, the model is more likely to output information that is inconsistent with actual needs or contaminated. However, the reasoning model can identify the errors, indicating that the model's reasoning ability is crucial in identifying the user's input intention.

Figure~\ref{fig:case5} shows that when faced with ambiguous or uncertain expressions, large models usually tend to customize questions or give default explanations based on their own training preferences, rather than actively asking users for more context.The GPT series performs better than models such as DeepSeek in such tasks, indicating that even models with strong reasoning capabilities may have difficulties in dealing with modal expressions such as sarcasm and irony. This reflects the limitations of AI in understanding complex human language expressions, as well as differences in the coverage of such language phenomena in the training data and the depth of the model's understanding of the social and cultural context. 

Figure~\ref{fig:case6} shows that for implicit intentions such as sarcasm, metaphors, and emotions, the model often only focuses on the literal meaning and has difficulty grasping the deep emotions or context, thereby outputting incorrect analysis results.

As shown in Table~\ref{tab:singlecol_split}, Figure~\ref{fig:case3} to Figure~\ref{fig:case6}, in the intention reasoning stage, all models encountered difficulties and performed poorly in “Inconsistent Instruction Reasoning”, “Fuzzy Language Interpretation” and “Intention Clarification Failure”; in “Misinformation Reasoning”, only GPT-o3 and Deepseek-R1 achieved good performance, while the other models underperformed. This suggests that all models have significant challenges in reasoning about intentions.

\subsection{Case of Reliable Dialog Generation} 
Current large language models exhibit both deterministic response preferences and ambiguous knowledge boundaries during generation. When queries fall within the model’s knowledge coverage, the outputs are generally reliable; however, in open-domain or previously unseen scenarios, the quality of generated content fluctuates significantly. 

Figure~\ref{fig:case7} illustrates the instability of LLM-generated content, which primarily manifests in two typical issues: First, when the model lacks relevant knowledge or information regarding the input (Q.1), it tends to fabricate details, producing content inconsistent with facts and thereby substantially compromising the accuracy and reliability of its outputs. Second, even when the model possesses relevant knowledge (Q.2), it may still misinterpret or improperly integrate contextual information, resulting in outputs that are contextually inappropriate or logically flawed. These issues not only undermine user experience but also limit the applicability of LLMs in high-stakes scenarios. 

Figure~\ref{fig:case8} demonstrates the misalignment between model-generated content and human values, mainly reflected in the model’s potential to violate widely accepted moral and ethical standards. On one hand, when handling sensitive or controversial topics, the model may generate harmful, offensive, or discriminatory statements, negatively impacting users. On the other hand, it may also produce responses that encourage illegal activities or contravene social ethics, introducing potential legal and societal risks and imposing higher requirements on the safety and trustworthiness of LLMs.

As shown in Table~\ref{tab:singlecol_split}, Figure~\ref{fig:case7} and Figure~\ref{fig:case8}, in the reliable dialog generation stage, all models failed to maintain “Response Stability”; in “Alignment” test, only GPT-o3 and Deepseek-R1 performed well, while the rest of the models failed. The results show that all models have clear shortcomings in “Response Stability”, and the reasoning models have improved in “Alignment”.

Overall, reasoning capabilities can improve the performance of llms for interacting with real human user commands, however, all models still face significant challenges and remain underpowered in real-world and human interaction scenarios.

\begin{table*}[h]
\small
\resizebox{\textwidth}{!}{
\begin{tabular}{clccccp{5.5cm}cc}
\hline
Cat. & Benchmark & Year & Lang. & Num. & Type & Description \\ \hline
\multirow{6}{*}{Instrution  Understanding} 
& BotChat~\cite{duan-etal-2024-botchat} & 2023 & En\&Zh & 7658 & M & Multi-round dialogue eval. via simulated data \\
& MINT~\cite{wang2024mint} & 2023 & En & 29,307 & M & Tool use and feedback in multi-turn dialogue \\

& MT-BENCH-101~\cite{bai-etal-2024-mt} & 2024 & En & 1,388 & M & Multi-turn dialogue ability \\

& $\infty$\{B\}ench~\cite{zhang-etal-2024-bench} & 2024 & En\&Zh & 100k & S & Long-context handling \\

& L-Eval~\cite{an-etal-2024-l} & 2024 & En & 2,000 & S & Evaluation of Long-Context Language Models \\

& LongICLBench~\cite{li2025longcontext} & 2025 & En & 2,100k & S & Long In-Context Learning \\

\hline

\multirow{4}{*}{Intention Reasoning} 
& BIPIA~\cite{yi2023benchmarking} & 2023 & En & 712.5K & S & Vulnerability to hint injection \\
& Miko~\cite{lu2024miko} & 2024 & En & 10k & S & Multimodal social intent understanding \\
& CONTRADOC~\cite{li2023contradoc} & 2023 & En & 891 & S & Self-contradiction in long docs \\
& CDCONV~\cite{zheng2022cdconv} & 2022 & Zh & 12K & M & Contradiction in Chinese dialogues \\ \hline

\multirow{3}{*}{Relibale Dialog Generate} 
& Open-LLM-Leaderboard~\cite{ye2024benchmarking} & 2024 & En & 10K & S & Uncertainty in generation \\
& ETHICS~\cite{hendrycks2020aligning} & 2020 & En & 130K & S & Moral reasoning \\
& FACTOR~\cite{muhlgay2023generating} & 2023 & En & 300 & S & Factuality in generated text \\ \hline
\end{tabular}
}
\caption{A selection of widely used benchmark datasets for evaluating LLMs. Th“Cat.”: task stage; 'Lang.': language of the benchmark, a; 'Num.': data size; 'Type': S=Single-round, M=Multi-round.}
\label{tab:benchmark}
\vspace{-6mm}
\end{table*}

\section{Benchmark}\label{sec:Benchmark}
This section covers benchmarks for LLMs in above three stage (\Cref{tab:benchmark}).

\subsection{Benchmarking Instruction Understanding}

Instruction understanding in LLMs involves extracting key information, maintaining coherence, and adapting to dynamic conversation changes, especially in long or multi-round dialogues. LLM capabilities in multi-turn dialogue and long-context processing have been explored through various benchmarks. BotChat~\cite{duan-etal-2024-botchat} evaluates dialogue generation, showing GPT-4's strengths but noting instruction compliance and length limitations in other models. MINT~\cite{wang2024mint} highlights limited progress in tool use and feedback for complex tasks. MT-Bench-101~\cite{bai-etal-2024-mt} identifies challenges in enhancing long-term interaction skills. For long-context tasks, performance drops in ultra-long texts (Zhang et al.~\cite{zhang-etal-2024-bench}), L-Eval~\cite{an-etal-2024-l} emphasizes length-instruction-enhanced metrics, and LongICLBench~\cite{li2025longcontext} reveals difficulties in reasoning across multiple pieces of information.

\subsection{Benchmarking LLM Reasoning}
LLM intention reasoning involves inferring user intentions by interpreting both explicit and implicit language cues.
The existing benchmarks comprehensively evaluate the multifaceted reasoning capabilities of LLMs, encompassing vulnerability to indirect hint injection attacks (BIPIA~\cite{yi2023benchmarking}), advantages in multimodal intention understanding (Miko~\cite{lu2024miko}), analysis of self-contradictions in long documents (CONTRADOC~\cite{li2023contradoc}), and contradiction detection in Chinese dialogues (CDCONV~\cite{zheng2022cdconv}). Collectively, these benchmarks highlight both the challenges and advancements of LLMs in complex reasoning.

\subsection{Benchmarking LLM Generation}
LLM Generation assesses a model's ability to understand user instructions, avoid fabricating false information, and generate accurate, contextually appropriate responses.
Open-LLM-Leaderboard~\cite{ye2024benchmarking}, ETHICS~\cite{hendrycks2020aligning} and FACTOR~\cite{muhlgay2023generating} all focus on the reliability evaluation of content generated by large models. Open-LLM-Leaderboard finds that large-scale models have higher uncertainty, and fine-tuned models have higher accuracy but greater uncertainty. ETHICS focuses on the ethical value alignment of generated content. FACTOR evaluates factuality through scalable methods to ensure that diverse and rare facts are covered. 

\section{Future Directions}\label{sec:Directions}
This section summarizes ongoing challenges in above stages with LLMs and outlines potential future research directions.

\paragraph{Automated Annotation Framework} Although LLMs excel in general-domain tasks, they often produce hallucinated or incomplete content in specialized fields due to limited domain-specific training data. While contextual learning and instruction fine-tuning methods have been explored to address this issue, manual data annotation remains labor-intensive and prone to quality inconsistencies. An automated annotation framework could streamline data labeling, enhancing model performance in specialized fields by ensuring higher quality and scalability of domain-specific training datasets.

\paragraph{GraphRAG} 
LLMs have shown impressive language generation capabilities through pre-training on large datasets, but their reliance on static data often results in inaccurate or fictional content, particularly in domain-specific tasks. The Graph-enhanced generation approach aims to tackle this by leveraging KGs and GNNs for precise knowledge retrieval. Despite its advantages, GraphRAG faces challenges in capturing structural information during graph reasoning tasks and struggles with multi-hop retrieval accuracy and conflict resolution between external and internal knowledge. Future work should focus on refining retrieval strategies and improving the stability and accuracy of GraphRAG in complex tasks.

\paragraph{Quantifying Uncertainty in LLMs} 
As LLMs like LLama and ChatGPT have revolutionized content generation, ensuring the reliability of their outputs remains a critical challenge. Uncertainty quantification is a promising approach to address this, enabling models to provide a confidence assessment alongside their responses. Evidence learning~\cite{sensoy2018evidential,amini2020deep}, an emerging method in uncertainty representation, offers a reliable approach for quantifying uncertainty directly from data. However, its application to LLMs is computationally intensive, and most existing work focuses on small-scale models. Future research should aim to optimize uncertainty quantification for large-scale models efficiently.

\paragraph{Balancing Safety and Performance} Although advancements in alignment techniques have improved factual accuracy and safety, they often come at the cost of the model's creativity and fluency. Striking a balance between safety and performance is crucial. Future research should explore new alignment methods that ensure both the safety and usability of LLMs, optimizing the trade-off between generating reliable, safe content and maintaining the model's creative and contextual capabilities.

\section{Conclusion}\label{sec:Conclusion}
This paper analyzes LLMs' performance in processing user instructions. Despite progress in natural language understanding, LLMs struggle with complex, inconsistent instructions, often resulting in biases, errors, and hallucinations. Improvements through prompt engineering, model expansion, and RLHF et al. have not fully addressed LLMs' limitations in reasoning and comprehension, limiting their real-world applicability.
We identify three challenges: instruction understanding, intention reasoning and reliable dialog generation. Future research should focus on enhancing reasoning for complex instructions and aligning outputs with user intention to improve LLMs' reliability.

\bibliographystyle{elsarticle-num}
\bibliography{elsarticle-template-harv}
\end{document}